\documentclass[conference]{IEEEtran} 
\usepackage{fancyhdr}
\usepackage{epsfig}
\usepackage{threeparttable}
\usepackage{epsf,epsfig}
\usepackage{amsthm}
\usepackage{amsmath}
\usepackage{amssymb}
\usepackage{amsfonts}
\usepackage[noadjust]{cite}
\usepackage{dsfont}
\usepackage{subfigure}
\usepackage{color}

\newtheorem{lemma}{Lemma}
\newtheorem{corollary}{Corollary}

\newtheorem{proposition}{Proposition}

\def\E{\mathsf{E}}

\def\l{\left}
\def\r{\right}
\def\({\left(}
\def\){\right)}

\def\b0{{\mathbf{0}}}

\begin{document}
%\IEEEoverridecommandlockouts
%\IEEEpubid{\makebox[\columnwidth]{978-1-4799-3086-9/14/\$31.00~\copyright~2014 IEEE \hfill} \hspace{\columnsep}\makebox[\columnwidth]{ }}

\title{Content-Specific Broadcast  Cellular Networks based on User Demand Prediction: A Revenue Perspective}

\author{\IEEEauthorblockN{Jihong Park and Seong-Lyun Kim}
\IEEEauthorblockA{Dept. of Electrical \& Electronic Engineering\\
Yonsei University\\
50 Yonsei-ro, Seodaemun-gu, Seoul, Korea\\
Email: \{jhpark.james, slkim\}@yonsei.ac.kr}
}

\maketitle

\begin{abstract}
The Long Term Evolution (LTE) broadcast is a promising solution to cope with exponentially increasing user traffic by broadcasting common user requests over the same frequency channels. In this paper, we propose a novel network framework provisioning broadcast and unicast services simultaneously. For each serving file to users, a cellular base station determines either to broadcast or unicast the file based on user demand prediction examining the file's content specific characteristics such as: file size, delay tolerance, price sensitivity. In a network operator's revenue maximization perspective while not inflicting any user payoff degradation, we jointly optimize resource allocation, pricing, and file scheduling. In accordance with the state of the art LTE specifications, the proposed network demonstrates up to $32$\% increase in revenue for a single cell and more than a $7$-fold increase for a $7$ cell coordinated LTE broadcast network, compared to the conventional unicast cellular networks.
\end{abstract}
\begin{IEEEkeywords}
LTE broadcast, eMBMS, unicast, resource allocation, delay, scheduling, pricing, revenue maximization
\end{IEEEkeywords}

\section{Introduction}
Explosive user traffic increase in spite of scarce wireless frequency-time resources is one of the most challenging issues for the future cellular system design \cite{CiscoMobileTrafficForecast:2013}. LTE broadcast, also known as evolved Multimedia Multicast Broadcast Service (eMBMS) in the Third Generation Partnership Project (3GPP) standards \cite{LecGab:EMBMSinLTERel11:Nov:2012}, is one promising way to resolve the problem by broadcasting common requests among users so that it can save frequency-time resources \cite{Ericsson:LTEBCRevEnab:2013}. The common user requests can be easily found in, for example, popular multimedia content or software updates in smart devices. By harnessing these overlapping requests of users, LTE broadcast enhances the total resource amount \emph{per cell}. This plays a complementary role to the prominent small cell deployment approach providing more resource amount \emph{per user} by means of reducing cell sizes \cite{Andrews:SvnWayHetNet:2013}.

To implement this technique in practice, it is important to validate the existence of sufficiently large number of common requests. According to the investigation in \cite{ChaKwkETAL:ITubeYouTube:2007}, discovering meaningful amount of common requests is viable even in YouTube despite its providing a huge amount of video files. That is because most users request popular files; for instance, 80\% of user traffic may occur from the top 10 popular files. On the basis of this reason, AT\&T and Verizon Wireless are planning to launch LTE broadcast in early 2014 to broadcast sports events to their subscribers \cite{FierceWireless:ATT700LTEBC:2013}.

The number of available common requests and its resultant saving amount of resources in cellular networks are investigated in \cite{SMYuSLKim:2013}, but it focuses on broadcast (BC) service while neglecting the effect of incumbent unicast (UC) service. Joint optimization of the resource allocations to BC and UC are covered in \cite{MsrCbgETAL:JointUCEMBMS:2010, LouQui:QoSSchRscEMBMS:2011} in the perspectives of average throughput and spectral efficiency. The authors however restrict their scenarios to streaming multimedia services where data are packetized, which cannot specify the content of data as well as the corresponding user demand of the files.

Leading from the preceding works, we propose a BC network framework being specifically aware of content and able to transmit generic files via either BC or UC service. The selection of the service depends on the following content characteristics: 1) file size, 2) delay tolerance, and 3) price discount on BC compared to UC. These characteristics are able to represent a content specified file in practice. For easier understanding, let us consider a movie file as an example. It is likely to be large file sized, delay tolerable (if initial playback buffer is saturated), and sensitive to the per-bit price of BC under usage-based pricing \cite{SenJoeETAL:IncTstDatSurv:2012} owing to its large file size. An update file of a user's favorite application in smart devices can be a different example, being likely to be small file sized, delay sensitive, and less price sensitive.

Furthermore, this study devises a policy that a base station (BS) solely carry out BC/UC service selection based on user demand prediction. Corresponding to the policy, we maximize the network operator's revenue without user payoff degradation by jointly optimizing BC resource allocation, file scheduling, and pricing. To be more specific, the following summarizes the novelty of the proposed network framework.
\begin{itemize}
\item \textbf{BC/UC selection policy}: a novel BC/UC selection policy is proposed where a BS solely assigns one of the services for each user by comparing his expected payoffs of BC and UC if assigned, without degrading user payoff.
\item \textbf{BC resource allocation}: optimal BC frequency allocation amount is derived in a closed form, showing the allocation is linearly increased with the number of users in a cell, and inversely proportional to UC price.
\item \textbf{BC pricing}: optimal BC price is derived in a closed form, proving the price is determined proportionally to the number of users until BC frequency allocation uses up the entire resources.
\item \textbf{BC file scheduling}: optimal BC file order is derived in an operation-applicable form as well as a closed form for a suboptimal rule suggesting smaller sized and/or more delay tolerable files should be prioritized for BC.
\end{itemize}

As a consequence, we are able to not only estimate revenue in a closed form, but also verify the revenue from the proposed network keeps increasing along with the number of users unlike the conventional UC only network where the revenue is saturated after exhausting entire frequency resources. Considering 3GPP Release 11 standards, we foresee up to $32$\% increase in revenue for a single LTE broadcast scenario and more than a $7$-fold increase for a multi-cell scenario.

%
%The remainder of the paper is organized as follows. The system model is described in Section II. The four network components above are optimized, and the resultant revenue is presented in Section III. The results are evaluated by simulation in Section IV, followed by concluding remarks in Section V.

\section{ System Model}

A single cellular BS simultaneously supports downlink UC and BC services with $W$ frequency bandwidth where BC files are slotted in a single queue. The BS serves $N$ number of mobile users who are uniformly distributed over the cell region. Let the subscript $k$ indicate the $k$-th user for $k \in \{1, 2, \cdots, N \}$, and define $\phi_k$'s as the locations of users. User locations are assumed to be fixed during $T$ time slots, but change at interval of $T$ independent of their previous locations. Let the subscripts $u$ and $b$ represent UC and BC hereafter, and $P_u$ and $P_b$ respectively denote UC and BC usage prices per bit. In order to promote BC use, the network offers price discount on BC so that it can compensate longer delay of BC.

\begin{figure}
\centering
	\includegraphics[width=7.7cm]{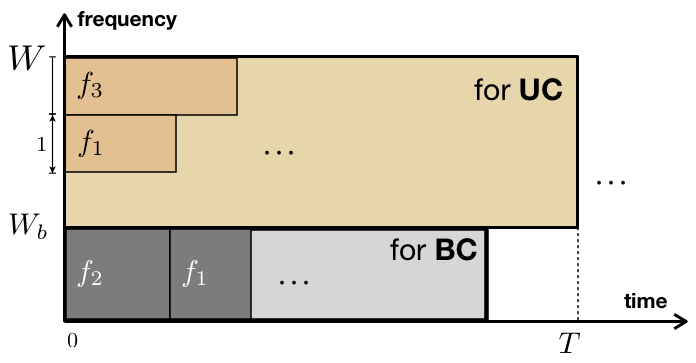}  
	\caption{Time-frequency resource allocation for unicast and broadcast services where $W_b$ amount of frequency is allocated for broadcast while unity is allocated for unicast during $T$ time slots} \label{Fig:RscAlloc}
\end{figure}

\subsection{User Request Pattern}
Each user independently requests a single file at the same moment with a unit interval $T$ time slots. Let the subscript $i$ represent the $i$-th popular file for $i\in \{1, 2, \cdots, M \}$ where $M$ denotes the number of all possible requests in a given region. Assume user request pattern follows Zipf's law (truncated discrete power law) as in YouTube traffic \cite{ChaKwkETAL:ITubeYouTube:2007}. It implies the file $i$ requesting probability $p_i$ is given as $i^{-\gamma}/H$ where $H = \sum_{j=1}^M j^{-\gamma}$ for $\gamma>0$. Note that larger $\gamma$ indicates user requests are more concentrated around a set of popular files.

\subsection{Network Operation} \label{Section:NetOperation}

The following example sequentially describes the BS's operation to serve a typical user $k$ requesting file $i$.

\begin{enumerate}
\item \textbf{Common request examination}: by inspecting user requests, BS becomes aware of the file $i$'s size $f_i$ as well as the number of file $i$ requests $n_i$.
\item \textbf{Delay tolerance examination}: user $k$ marks his requesting priority of the file $i$ as in conventional peer-to-peer (P2P) services (e.g. high/low). Assuming BS has the full knowledge of users' quality-of-experience (QoE) patterns, this priority information corresponds to delay threshold $\theta_{ik}$, allowable delay without degrading QoE. 
\item \textbf{BC frequency allocation, pricing, and file scheduling}: by inspecting $f_i$, $n_i$, and $\theta_{ik}$, BS allocates BC frequency amount $W_b$, and sets BC price $P_b$ as well as optimizing BC file scheduling in a revenue maximizing order.
\item \textbf{BC/UC selection}: meanwhile in 3), BS assigns either BC or UC to user $k$ in order to maximize revenue without inflicting the user's payoff loss. %We neglect any delay through 1-4) without loss of generality.
\end{enumerate}

Note that the pricing scheme we consider is similar to time-dependent pricing \cite{SenJoeETAL:IncTstDatSurv:2012} in respect of its flattening user traffic effect by adjusting $P_b$ over time. The target offloading traffic by the pricing is, however, novel since the conventional scheme aims at the entire user traffic but the proposed at \emph{content-specific} traffic captured by $n_i$. %This will be further discussed in Section \ref{Section:RevMax}. 

\subsection{Resource Allocation}

BS allocates $W_b$ amount of BC frequency for handling the entire BC assigned requests. In compliance with the 3GPP Release 11 \cite{LecGab:EMBMSinLTERel11:Nov:2012}, the earmarked amount cannot be reallocated to UC requests during $T$ as Fig. \ref{Fig:RscAlloc} visualizes. For each UC request, BS allocates a normalized unity frequency resource, to be addressed with a realistic unit in Section \ref{Section:Num}.

\begin{figure}
\centering
	\includegraphics[width=4.1cm]{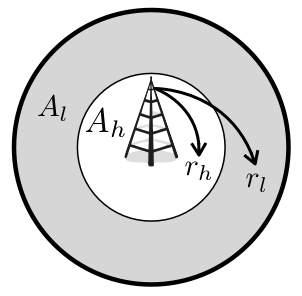}  
\caption{Wireless channel model where a cellular base station provides average rate $r_h$ for region $A_h$, and $r_l$ for $A_l$} \label{Fig:Channel}
\end{figure}

\subsection{User Payoff} \label{SubSection:UserPayoff}
Let $U_{ik}$ denote the payoff of user $k$ when downloading file $i$ via UC. Consider the payoff has the following characteristics: logarithmically increasing with $f_i$; logarithmically decreasing with its downloading completion delay after exceeding $\theta_{ik}$ \cite{Shaikh:QoEUsrNet:2010}; and linearly decreasing with cost under usage-based pricing \cite{SenJoeETAL:IncTstDatSurv:2012}. Define $r_k^u$ as the spectral efficiency when user $k$ is served by UC. Consider delay sensitive UC users such that UC downloading completion delays always make them experience QoE degrading delays.
, i.e. $f_i/r_k^u > (\theta_{ik} + 1)$. Additionally, we neglect any queueing delays on UC. The payoff $U_{ik}$ then can be represented as follows.
 \begin{equation} \small
U_{ik} = \log\l( \frac{1 + f_i}{ f_i /r_k^u - \theta_{ik}}\r) -P_u f_i 
\end{equation} \normalsize

Note that $U_{ik} >0$ as we are only interested in the users willing to pay for at least UC service.

In a similar manner, consider $B_{ik}$ indicating the payoff of user $k$ when downloading file $i$ via BC. Let $r_k^b$ denote the BC spectral efficiency of user $k$. We further define $s_i$ as the size of the broadcasted files until the BC downloading of file $i$ completes. This captures the effect of BC file scheduling. The payoff $B_{ik}$ can be represented as below.
% Equation:
\begin{equation} \small
B_{ik} = \ \log\l( \frac{1 + f_i}{s_i / \l(W_b r_k^b\r) -\theta_{ik}} \r) -P_b f_i \;.
\end{equation} \normalsize

To maximize revenue while guaranteeing at least UC payoff amount, BS compares $U_{ik}$ and $B_{ik}$, and assigns either UC or BC service, to be further elaborated in Section \ref{Section:RevMax_A}.

\subsection{Wireless Channel} \label{Section:Channel}
We consider distance attenuation from difference user locations $\phi_{k}$, and adaptive modulation and coding (AMC) which changes modulation and coding schemes (MCS) depending on wireless channel quality \cite{RdhkTirETAL:AMCMBSFN:2012}. While UC can adaptively adjust MCS based on its serving user's channel quality, the MCS for BC resorts to aim at the worst channel quality user because BC has to apply an identical MCS to all its users. BC average spectral efficiency is therefore not greater than the UC's.

To be more specific, as Fig. \ref{Fig:Channel} illustrates, we consider a cell region $A$ divided into $A_h$ and $A_l$. BS can provide high spectral efficiency $r_h$ to $A_h$, but low spectral efficiency $r_l$ to $A_l$ for $r_l \leq r_h$. Let $|A|$ denote the area of a region $A$. The probability that user $k$ is located within $A_h$, $\Pr\l\{ \phi_k \in A_h \r\}$, is given as $|A_h|/ |A|$, independent of $k$ \cite{StoyanBook:StochasticGeometry:1995}. Define $r_u$ as UC average spectral efficiency of user $k$, represented as:
\begin{equation} \small
r_u = r_l+ (r_h - r_l) \Pr\l\{ \phi_k \in A_h \r\}.
\end{equation}

Similarly, average BC spectral efficiency $r_b$ is given as:
\small\begin{eqnarray} 
r_b &=& r_l + (r_h - r_l) \Pr\l\{ \phi_k \in A_h \r\}^{N_b} \\
&\approx& r_l \qquad \text{as $N \rightarrow \infty$  } \label{Eq:r_b}
\end{eqnarray} \normalsize
where $N_b$ denotes the number of BC users. Note that \eqref{Eq:r_b} is because $N_b$ is an increasing function of $N$.

\section{Revenue Maximizing BC Network Management} \label{Section:RevMax} 

In order to maximize revenue, we optimize BC frequency bandwidth $W_b$, price $P_b$, and file scheduling. For more brevity, assume sufficiently large $N$ such that BC average spectral efficiency is approximated as $r_l$ as in \eqref{Eq:r_b}.

\subsection{BC/UC Selection Policy and Problem Formulation} \label{Section:RevMax_A}
We firstly propose a BC/UC selection policy guaranteeing allowable user payoff, and then formulate the average revenue maximization problem under the policy. Assume that users predict to be served by UC as default, and hence BS should guarantee at least the amount of UC payoff for every service selection. For user $k$, revenue maximizing service selection policy is described in the following two different user payoff cases:
\begin{enumerate}
\item If $B_{ik} \geq U_{ik}$, BS firstly assigns UC as much as possible until UC resource allocation reaches $(W-W_b)T$ because $P_u \geq P_b$. After using up the entire UC resources, BS then assigns BC;
\item If $B_{ik} < U_{ik}$, BS resorts to assign UC in order to avoid payoff loss.
\end{enumerate}

Note that this policy not only maximizes revenue, but also, albeit not maximizes, enhances user payoff.

For simplicity without loss of generality, assume the required resource amount for UC user demand exceeds the entire UC resources, $(W - W_b)T$. As there is no more available UC resource, $P_u$ is set as a maximum value due to no price discount motivation on UC. It results in the revenue from UC is fixed as $P_u(W - W_b)T$. By contrast, the revenue from BC still can be increased if $B_{ik} \geq U_{ik}$ holds. As a consequence, the average revenue in a cell region $A$ is represented as follows.
%i.e. $\sum_{i=1}^M f_i \sum_{k=1}^{n_i} \textbf{1}\l\{ U_{ik}>0\r\} \geq r_u WT$. 
% Equation:
\begin{equation*}  \small
\mathcal{L}_0:= \E_k \bigg[ P_b \sum\limits_{i=1}^M f_i \sum\limits_{k=1}^{n_i}  \textbf{1}\l\{ B_{ik}  \geq U_{ik} \r\} \bigg]  + P_u \l( W - W_b\r)T
\end{equation*} \normalsize
The left and right halves of $\mathcal{L}_0$ respectively indicate the average revenues from BC and UC, and $\textbf{1}\l\{ \cdot \r\}$ is an indicator function which becomes 1 if a condition inside the function is satisfied, otherwise 0. Unfortunately, $\mathcal{L}_0$ is an analytically intractable nonlinear function due to $\textbf{1}\l\{ B_{ik} \geq U_{ik}\r\}$. In order to detour the problem, consider the following Lemma.

%__Lemma: 
%
\begin{lemma}\emph{
For $(P_u - P_b)f_i <1$, the inequality $\mathcal{L}_0 \geq \mathcal{L}$ holds where $\mathcal{L}$ is defined as:
% Equation:
\small\begin{eqnarray*}
P_b N  \sum_{i=1}^M f_i p_i  \l[1 - \frac{s_i  \theta_i r_u}{W_b r_b} \l\{ 1 - \l(P_u- P_b\r)f_i \r\}   \r] + P_u(W - W_b)T
\end{eqnarray*} \normalsize
and $\theta_i : = \E_k\l[ 1/\l( f_i - r_k^u \theta_{ik}\r) \r]$.\\
%______Proof of Lemma
%
\noindent\begin{proof}
See Appendix.
\end{proof}
%
%______Proof of Lemma ends
}\end{lemma}
\vspace{-5pt}
%
%__Lemma Ends
Note that $\theta_i$ indicates the aggregate delay tolerance of file $i$ among users for a given $f_i$ and $r_k^u$. Additionally, the assumption $(P_u - P_b)f_i<1$ does not imply small sized files since $f_i$ is a normalized value. Applying $\mathcal{L}$ in the result of Lemma 1, the lower bound of $\mathcal{L}_0$, yields the corresponding problem formulation given as:
\begin{align}\small
      \textbf{P1}.&
   \begin{aligned}[t]
    & \underset{W_b,  P_b, s_i  }{\text{max}} \; \mathcal{L}  \notag\\
   \end{aligned}  \label{Eq:Problem1}  \notag\\
   &\text{subject to} \notag \\
   & \qquad \quad  0\leq P_b \leq P_u, \notag \\
   & \qquad \quad   0 \leq W_b \leq W, \notag \\
   & \qquad \quad   s_i > s_j \; \text{or} \; s_i<s_j, \; \forall i, j \in \l\{ 1, 2, \cdots, M\r\}. \notag
\end{align} \normalsize
\noindent The last inequality condition means BC files are slotted in a single queue while BS transmits each file only once. In respect to $\mathcal{L}$ in $\textbf{P1}$, the following sections sequentially derive optimal BC network components, $W_b^*$, $P_b^*$, and $s_i^*$.

%\begin{align}
%   &
%   \begin{aligned}[t]
%   & \underset{s_i,  P_b,  W_b }{\text{max}}  \; P_b \sum\limits_{i=1}^M f_i \sum\limits_{k=1}^{n_i} \E_t \l[ 1 - e^{-\l(\pi_{ik}^b - \pi_{ik}^u \r) } \r] \\
%   & \qquad \qquad + P_u \l( W - W_b\r)T 
%   \end{aligned} 
%\end{align} 

\subsection{BC Frequency Allocation} \label{Section:W_b}
Define $F$ as $ \sum_{i=1}^M f_i p_i$ implying the average requesting file size per user, which is a given value independent of our network design.  Consider small $f_i$ and sufficiently large $N$  as assumed at the beginning of Section \ref{Section:RevMax}, we can derive a closed form solution of the optimal BC frequency allocation in the following Proposition.

%__Proposition: 
%
\begin{proposition}\emph{
Optimal BC frequency allocation $W_b^*$ is given as follows.
\small \begin{equation*}
W_b^* \approx \min\l( \frac{N F}{4 P_u T}, W \r)
\end{equation*}\normalsize
%______Proof of Proposition
%
\noindent\begin{proof}
See Appendix.
\end{proof} \vspace{-5pt}
%
%______Proof of Proposition ends
}\end{proposition}
%
%__Proposition end
The proposition shows the optimal BC frequency allocation is determined regardless of BC spectral efficiency $r_b$ and price $P_b$. Moreover, it provides the network design principles that the BC frequency amount is proportional to $N$ and inversely proportional to UC price $P_u$. The latter is because it becomes necessary to enhance BC downloading rate by allocating more amount of frequency to BC when BC service becomes less price competitive (smaller $P_u$). 

%Secondly, insufficient wireless resource $T$ promotes to exploit BC service because of its thrifty nature in  frequency.

%
%\noindent \emph{Remarks}: 
%\begin{itemize}
%\item $W_b^*$ increases along with the following factors. \begin{itemize}
%\item $N\; \uparrow\; $; since $F$ linearly increases with $N$.
%\item $P_u \downarrow\; $; BC delay performance should be enhanced by allocating more $W_b$ when $P_b$ becomes less price competitive, small $(P_u - P_b)$ in other words.
%\item $T \; \downarrow \;$; for more insufficient wireless resources, BC necessity arises.
%\end{itemize}
%\item $W_b^*$ is independent of $r_b$ and $P_b$.
%\end{itemize}

\subsection{BC Pricing}
We can derive the optimal BC price in a closed form in the following Proposition.
%__Proposition: 
%
\begin{proposition}\emph{
Optimal BC price is given as follows.
\small \begin{equation*}
P_b^* \approx  \min \l\{ \frac{1}{2}\l(  \frac{N r_b F^2}{4 P_u T r_u S^*} + P_u   \r) , P_u \r\}
\end{equation*}\normalsize
where $S^* = \sum_{i=1}^M s_i^* \theta_i f_i p_i $\\ 
%______Proof of Proposition
%
\noindent\begin{proof}
See Appendix.
\end{proof}
%
%______Proof of Proposition ends
}\end{proposition} \vspace{-5pt}
%
%__Proposition end
The result shows that $P_b^*$ is strictly increasing with $N$ within the range from $P_u/2$ to $P_u$. It implies price increase is more effective to enhance revenue than price discount although the discount may promote more BC use. This result plays a key role to design a BC file scheduler for detouring a recursion problem in Section \ref{Section:Schedule}. In addition, it is worth mentioning that BC file scheduler affects $P_b^*$ by adjusting $S^*$ since $s_i^*$ therein varies along with the order of BC files, to be further elaborated in the following section.

%
%\begin{figure}
%\centering
%	\includegraphics[width=8.5cm]{Figs/Fig_order.eps}  
%	\caption{BC file scheduling convergence} \label{Fig:order}
%\end{figure}

\subsection{BC File Scheduler} \label{Section:Schedule}

Each file $i$ is tagged with a weighting factor $w_i$ by BS. BS examines the scheduling file priorities by comparing $w_i$'s. The file scheduling affects $s_i$ defined in Section \ref{SubSection:UserPayoff}, so we maximize $\mathcal{L}$ in terms of $s_i$ as follows.

\begin{proposition} (Optimal Scheduler) \emph{ Broadcasting files in a descending order of $w_i^*$ is the optimal scheduling rule maximizing $\mathcal{L}$ in $\textbf{P1}$ where
 \begin{equation*}\small
w_i^* := \theta_i p_i \l\{ 1 - \frac{f_i}{2} \l(P_u - \frac{N r_b F^2}{4 P_u T r_u S^*} \r) \r\}.
\end{equation*} \normalsize
% Equation:
%______Proof of Proposition
%
\noindent\begin{proof}
For a given $P_b^*$, consider the subproblem of \textbf{P1}:
\small\begin{align}
   \textbf{P2}. &
   \begin{aligned}[t]
   &\underset{s_i}{\text{min}}  \; \sum\limits_{i=1}^M  s_i \theta_i f_i p_i  \l\{ 1 - (P_u - P_b^*)f_i\r\}  \notag\\
   \end{aligned}  \label{Eq:Problem1}  \\
   &\text{subject to} \notag \\
   & s_i > s_j \; \text{or} \; s_i<s_j, \; \forall i, j\leq N. \notag
\end{align}  \normalsize
Applying the Smith's indexing rule in \cite{BertsimasBook:IntroLinearOpt:1997} and Proposition 2 leads to yield the result of the statement in Proposition 3.
\end{proof}
%
%______Proof of Proposition ends
}\end{proposition} \vspace{-5pt}
%The implication of the result is that scheduler design becomes less effective 
%if $N$ becomes larger, file size is less effective in the scheduler design.
Note that $w_i^*$ is recursive since $S^*$ in $w_i^*$ is a function of $s_i^*$ which is also a function of $w_i^*$. This cannot be solved analytically, and therefore we resort to derive the value by simulation in Section \ref{Section:Num}. In order to provide more fundamentally intuitive understanding, we consider the following suboptimal but closed form solution.
\begin{corollary} (Suboptimal Scheduler)\emph{ Broadcasting files in a descending order of $\bar{w}_i^*$ is a suboptimal scheduling rule enhancing $\mathcal{L}$ in \textbf{P1} where
\begin{equation*}
\bar{w}_i^* := \theta_i p_i \l( 1 - \frac{P_u f_i}{2} \r).
\end{equation*}
% Equation:
%______Proof of Corollary
%
\noindent\begin{proof}
Exploiting the boundary values of $P_b^*$ in Proposition 2 at Proposition 3 enables to bypass the recursion problem, completing the proof.
\end{proof}
%
%______Proof of Corollary ends
}\end{corollary}\vspace{-5pt}

%__Corollary Ends

\begin{figure}
\centering
	\subfigure[Scenario 1: Single cell LTE broadcast]{\includegraphics[width=9.48cm]{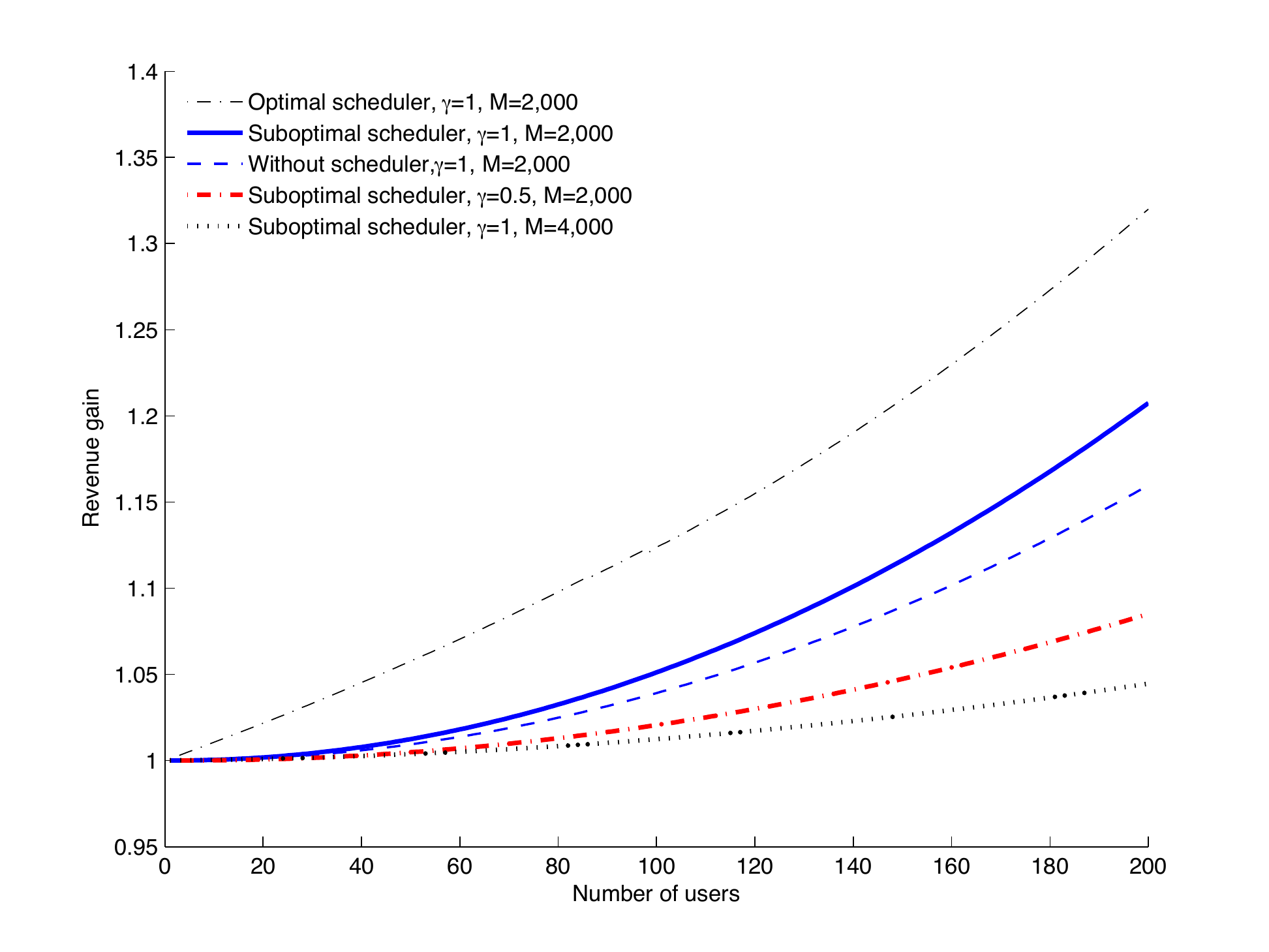}\label{Fig:R}} 
	\subfigure[Scenario 2: 7 cell coordinated LTE broadcast]{\includegraphics[width=9.48cm]{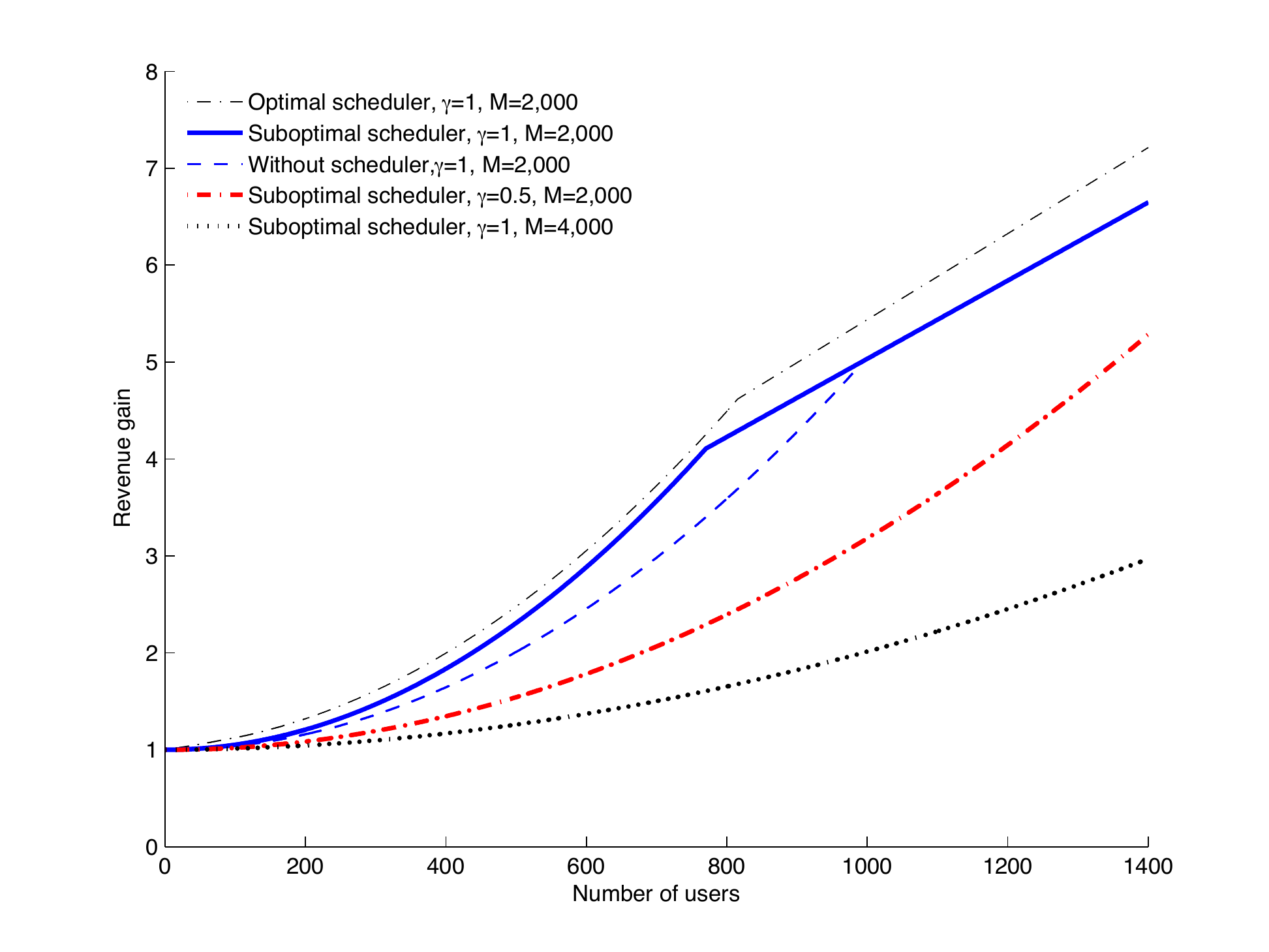}\label{Fig:R_MBSFN}}
	\caption{Revenue gains of (a) a single cell and (b) 7 cell coordinated LTE broadcast networks under the following environmtnets: with the optimal/suboptimal scheduler, without scheduler, lower popular file concentration $\gamma$ of user requests, and larger number of possible requesting files $M$}  \label{Fig:Rs}
\end{figure}

 Although the proposed scheduler is suboptimal, it still shows close-to-optimal behavior, to be verified by Fig. \ref{Fig:Rs} in Section \ref{Section:Num}. The suboptimal scheduler provides the following network design principle: more delay tolerable (larger $\theta_i$), more popular (larger $p_i$), and/or smaller files (smaller $f_i$) should be prioritized for BC if $f_i$ is sufficiently small such that $P_u f_i/2 < 1$.

\subsection{Revenue Gain} \label{Section:RevMax_R}

In a revenue perspective, we compare the proposed BC/UC network and conventional cellular networks where only UC operates. As a performance metric, we consider \emph{revenue gain} $R$ defined as the revenue of the proposed BC/UC network divided by that of the UC only network. By combining Propositions 1--3, our proposed network framework shows the following revenue gain.
%__Proposition: 
%
\begin{proposition}\emph{
The revenue gain $R$ is given as follows.
\small \begin{equation*}
R \approx 1 + \frac{NF}{2WT}   \l\{
             \min\l( \frac{N r_{b} F^2}{4 {P_{u}}^2 T r_u S^{*}}, 1 \r) + 1- \frac{G}{ P_u}   \r\}
\end{equation*} \normalsize
where $G:=  ( 0.5 + \sum_{i=1}^M s_i ^* \theta_i p_i  /S^* ) $\\
%______Proof of Proposition
%
\noindent\begin{proof}
Applying the results of Propositions 1--3 into $\mathcal{L}$ yields the following maximized revenue of the proposed network: $NF \l(   P_b^* - G/2 \r) + P_u W T$. Dividing it by the UC only network's revenue $P_u W T$ while applying Proposition 2 concludes the proof.
\end{proof}
%
%______Proof of Proposition ends
}\end{proposition} \vspace{-5pt}
%
%__Proposition end

Interestingly, the proposed network always achieves positive revenue gain for sufficiently large files such that $P_u>G$ where $G$ defined in Proposition 4 is a decreasing function of $f_i$ (recall $S^*$ in $G$ and $s^*$ therein is an increasing function of $f_i$ by definition in Section \ref{SubSection:UserPayoff}). For those files, the revenue gain $R$ increases with the order of $N^2$, converging to the order of $N$ for large $N$ when $P_b^* = P_u$ as the effect of $N$ diminishes. It is worth mentioning that $R$ grows even when frequency-time resources become scarce (smaller $WT$) thanks to the thrifty nature of BC in  frequency. In addition, the result captures the design of BC file scheduler affects revenue by adjusting $S^*$ (and G, a function of $S^*$).

% BC scheduler design can further enhance the gain by minimizing $S^*$. 

% The principle can explain the effect of file scheduler in revenue as follows. According to the principle, BS broadcasts a file having smaller size $f_i$, larger $\theta_i$ and $p_i$ ahead of the other files. Multiplying these three values by $s_i^*$, which is an increasing function of $f_i$, may decrease $S^*$ in Proposition 2 compared to a random file scheduler; for easier understanding, you can imagine a scheduler where larger sized files are prioritized for BC as an counter example, which increases $S^*$ compared to the proposed scheduler's.

%scheduled files according to the principle decreases $S^*$ defined in Proposition 2, which is an increasing function of $f_i$ by definition in Section \ref{SubSection:UserPayoff}).
 
%  For larger $\theta_i$'s and/or $p_i$'s, smaller files are scheduled ahead of the other files, and it decreases $S^*$ defined in Proposition 2 (recall $s_i^*$ in $S^*$ is an increasing function of $f_i$ by definition in Section . This corresponds to increasing $P_b^*$ in Proposition 2, resulting in more revenue.

\begin{figure}	
\centering
	\includegraphics[width=8.9cm]{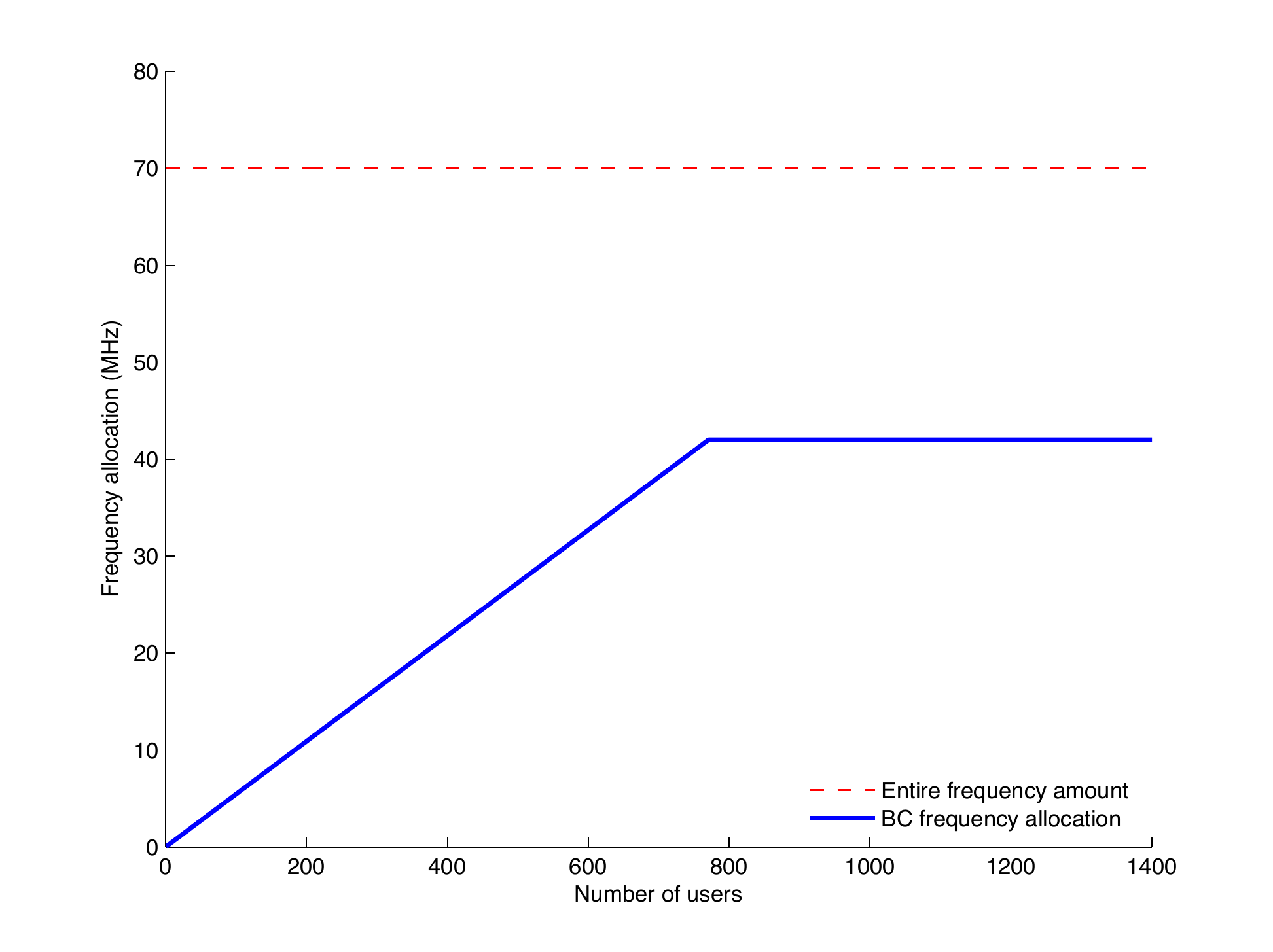}  
	\caption{Optimal broadcast frequency allocation of the 7 cell coordinated LTE broadcast network with the proposed suboptimal scheduler for increasing the number of users when $\gamma =1$ and $M=2$,$000$}  \label{Fig:W_b}
\end{figure}

\begin{figure}
\centering
	\includegraphics[width=8.9cm]{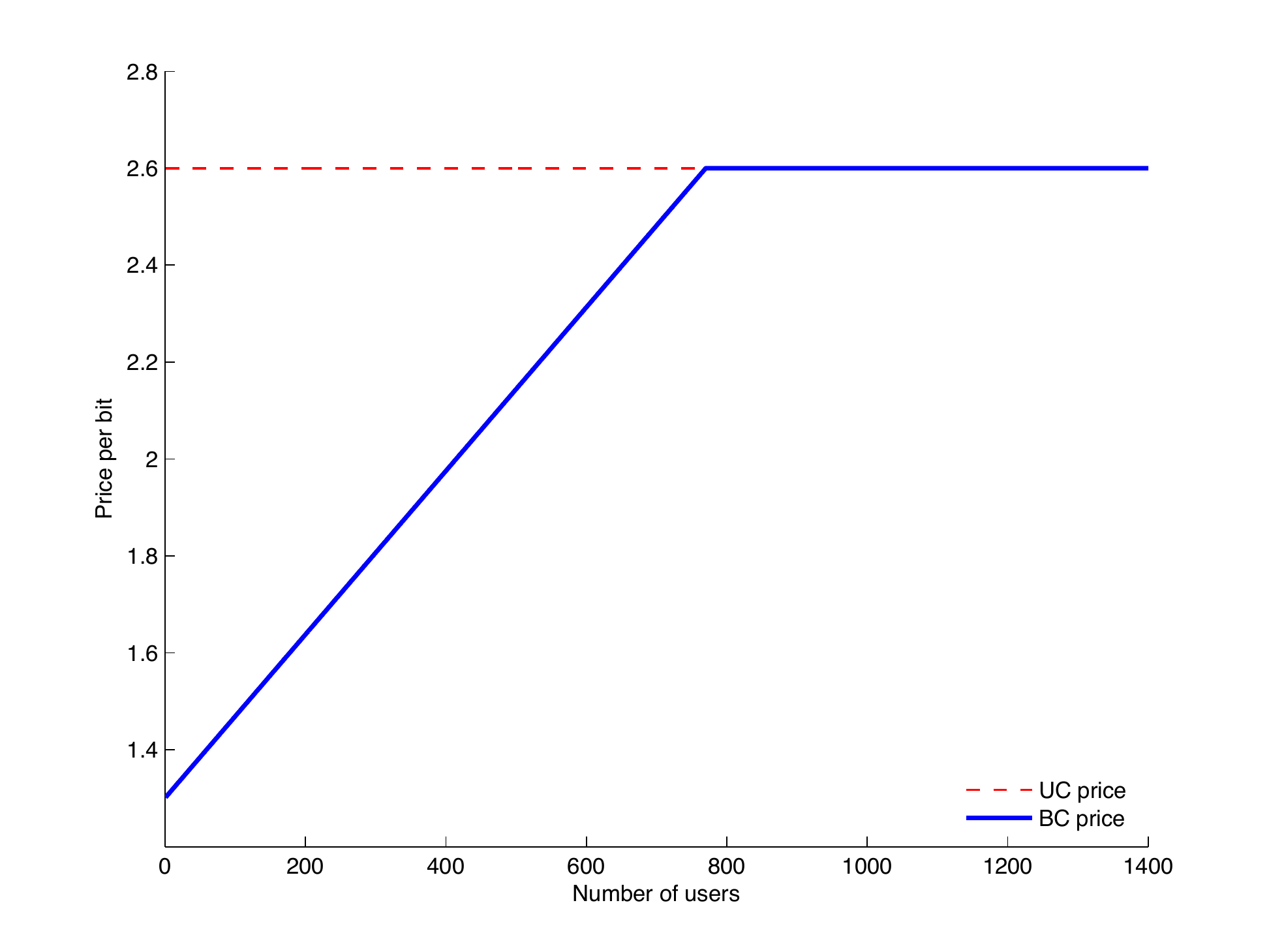}  
	\caption{Optimal broadcast price of the 7 cell coordinated LTE broadcast network with the proposed suboptimal scheduler for increasing the number of users when $\gamma =1$ and $M=2$,$000$} \label{Fig:P_b}
\end{figure}
\section{Numerical Results} \label{Section:Num}

We consider two different LTE broadcast network scenarios in accordance with 3GPP Release 11 standards \cite{LecGab:EMBMSinLTERel11:Nov:2012}. 

\subsection{Single Cell LTE Broadcast}
The first scenario is a typical single cell operates LTE BC, having the number of users $N$ up to $200$ with the entire frequency amount $W$ given as $10$ MHz. For BC, BS is able to allocate up to $60$\% of $W$. For UC, BS allocates average $2.5$ MHz to a single UC user until the downloading completes. At $A_h$, the average spectral efficiency $r_h$ is given as $2.4$ bps/Hz whereas $r_l$ at $A_l$ is 45 \% degraded from $r_h$ where $|A_l| = 9|A_h|$. These correspond to MCS index $19$ with 64QAM and the index $12$ with 16QAM respectively \cite{RdhkTirETAL:AMCMBSFN:2012}. The number of possible requesting files $M$ in the cell is fixed as $2$,$000$, and the Zipf's law exponent $\gamma$ is set as $1$ as default. File sizes are uniformly distributed from $160$ to $634$ MBytes, which may correspond with $4.8$ to $19$ minute long 1080p resolution video content. User delay threshold $\theta_{ik}$ is uniformly distributed from $0.6$ to $6$ seconds. Furthermore, $T$ is set as $2$ minutes and $P_u$ as $2.6$ a normalized value having no unit.

Fig. \ref{Fig:R} shows up to $32$\% gain in revenue for a single cell LTE broadcast network, including the effect of the $4.7$\% increment from the suboptimal scheduler proposed in Section \ref{Section:Schedule}. Moreover, scheduler design becomes more important when $N$ increases due to its increasing effect on revenue gain. In addition, the result captures the revenue gain is highly depending on user request concentration $\gamma$ (Zipf's law exponent) as well as the number of possible requesting file $M$ in a cell. Specifically, doubling $\gamma$ from $0.5$ decreases revenue gain by up to $12.7$\%, and $M$ from $2$,$000$ does by $16.3$\%.

\subsection{7 Cell Coordinated LTE Broadcast}
The second scenario we consider is a Multicast Broadcast Single Frequency Network (MBSFN) \cite{RdhkTirETAL:AMCMBSFN:2012} where $7$ neighboring cells are synchronized and operate LTE broadcast like a single cell. Assuming we neglect inter-cell interference, all the simulation settings are the same as in the single cell case except for the increased entire frequency amount $W$ by $70$ MHz and the number of users $N$ by up to $1$,$400$. As a result, Fig. \ref{Fig:Rs} shows the proposed network with the suboptimal scheduler achieves up to $720$\% revenue. The result also verifies that the revenue gain increasing rate with respect to $N$ converges to a linear scaling law when $P_b^* = P_u$ (see Fig. \ref{Fig:P_b} at $N \geq 770$) as expected in Section \ref{Section:RevMax_R} The effect of gain increment by the scheduler increases as anticipated in the single cell case for small $N$. This tendency, however, is no longer valid after exceeding $N=770$, where having the maximum $70.6$\% revenue increment by means of the suboptimal scheduler, and the effect of scheduler diminishes along with increasing $N$. The reason is there is no more available BC frequency since then, and thus revenue cannot be increased by any operations of BS other than the increasing number of common requests due to $N$. This behavior can be further justified by Fig. \ref{Fig:W_b} and \ref{Fig:P_b} respectively representing the linear growing rates of $W_b^*$ and $P_b^*$ with increasing $N$, as well as the convergence to the maximum values for $N\geq 770$.

\vspace{-5pt}

\section{Conclusion}

In this paper, we propose a BC network framework adaptively assigning BC or UC based on user demand prediction by examining content specific information such as file size, delay tolerance, and price sensitivity. For the purpose of the network operator's revenue maximization, the proposed framework jointly optimizes resource allocation, pricing, and file scheduling under a novel BC/UC selection policy. 

Although a BS solely assigns BC or UC service without informing users of the possible selections, the proposed policy does not degrade but even enhance user payoff. In addition, this study provides closed form solutions that enables to understand the fundamental behavior of the proposed framework and give meaningful network design insights; for instance, revenue gain scaling order becomes $N$ from $N^2$ as $N$ increases. We consequently observe up to $32$\% increase in revenue for a single cell and more than 7 times for 7 cell coordinated LTE broadcast networks compared to the conventional networks. 

The future work we are heading in is to extend the proposed framework into more general multi-cell scenarios which may rigorously incorporate inter-cell interference modeling.

\vspace{-5pt}

%______Ack._________________________
\section*{Acknowledgement}\small
This research was supported by the Ministry of Science, ICT and Future Planning, Korea, under the Communications Policy Research Center support program supervised by the Korea Communications Agency (KCA-2013-001). \normalsize
%This research was funded by the MSIP (Ministry of Science, ICT \& Future Planning), Korea in the ICT R\&D Program 2013
%\pagebreak
%\vspace{5cm}

\vspace{-5pt}

%by replacing the function by $1- e^{-C \max \l\{\pi_{ik}^b - \pi_{ik}^u,0 \r\}}$ for a constant $C>0$. For large $N$ and small $f_i <1$

\appendix
\emph{Proof of Lemma 1:} Let $X_k$ denote $\textbf{1}\l\{ B_{ik} >U_{ik} \r\}$. Since $X_k$'s are independent of $n_i$, we can apply Wald's identity \cite{GallagerBook:StochasticProcs:1995}, yielding $\E_k \l[ \sum_{k=1}^{n_i} X_k\r] = Np_i \E_k \l[ X_k\r]$. The lower bound of $X_k$ is derived as follows.
\begin{eqnarray}  \small
X_k &\geq & 1 - e^{ - \l( B_{ik}- U_{ik} \r)}  \\
%&=& 1 - \l( \frac{ s_i/(W_b r_k^b )- t_{ik}}{ f_i/r_k^u - t_{ik}}  \r) e^{-(P_u - P_b)f_i}\\
&\approx& 1 - \l( \frac{ s_i/(W_b r_k^b )- t_{ik}}{ f_i/r_k^u - t_{ik}}  \r) \l\{ 1 - (P_u - P_b)f_i \r\} \\
&\geq & 1 -  \frac{ s_i}{ W_b r_k^b \l( f_i/r_k^u - t_{ik} \r) }  \l\{ 1 - (P_u - P_b)f_i \r\} \label{Eq:indicatorLB}
\end{eqnarray} \normalsize
Combining these results completes the proof.
\hfill $\blacksquare$\\

\vspace{-5pt}
\emph{Proof of Proposition 1 and 2:}
The lower bound of average revenue gain $\mathcal{L}$ is a concave function with respect to $P_b$ as well as $W_b$. We therefore can find the unique optimal point $(P_b^*, W_b^*)$ via convex programming. Let $P_b$ be fixed, and consider $\mathcal{L}$ in terms of $W_b$, yielding the solution given as:
\begin{equation} \small
W_b^* = \sqrt{ \frac{P_b N r_u \sum_{i=1}^M s_i \theta_i {f_i}^2 p_i}{P_u T r_b}}. \label{Eq:W_b_pre}
\end{equation} \normalsize

Similarly, for a fixed $W_b$, the optimal BC price is given as follows.
\vspace{-5pt}
\begin{equation} \small
P_b^* = \frac{P_u}{2} + \l(4 \sum\limits_{i=1}^M s_i \theta_i {f_i}^2 p_i \r)^{-1} \sum\limits_{i=1}^M p_i \l( \frac{W_b}{r_u}  f_i -  s_i \theta_i \r) \label{Eq:p_b_pre}
\end{equation} \normalsize
Combining \eqref{Eq:W_b_pre} and \eqref{Eq:p_b_pre} proves Proposition 1. For Proposition 2, $N/S^*$ increases with $N$ since $s_i^* < N$ due to $f_i < 1$ where $s_i^*$ is only a function of $N$ in $S^*$. This proves $P_b^*$ is an increasing function of $N$, completing the proof.\hfill $\blacksquare$
\vspace{-5pt}

%%______Ack._________________________
%\section*{Acknowledgement}
%This work was supported by the Students' Association of the  Graduate School of Yonsei University funded by the Graduate School of Yonsei University.\\

%
%%______Ref._________________________
\bibliographystyle{ieeetr}

%\bibliography{IEEEabrv,LTEBC}

%\end{IEEEbiography}

\end{document}